\documentstyle[manuscript,aps,tighten]{revtex}
\begin{document}
\draft
\preprint{WUHEP/98-19}

\title{Massless quantum electrodynamics with a critical point}

\author{Carl M. Bender\thanks{Electronic address: cmb@howdy.wustl.edu}}
\address{Department of Physics, Washington University, St. Louis, MO 63130, USA}

\author{Kimball A. Milton\thanks{Electronic address: milton@mail.nhn.ou.edu}}
\address{Department of Physics and Astronomy, University of Oklahoma, Norman, OK
73019, USA}

\date{\today}
\maketitle

\begin{abstract}
Recently, it has been observed that a quantum field theory need not be Hermitian
to have a real, positive spectrum. What seems to be required is symmetry under
combined parity and time-reversal transformations. This idea is extended to
massless electrodynamics, in which the photon couples to the axial-vector
current with an imaginary coupling constant. The eigenvalue condition necessary
for the finiteness of the theory can now be solved; the value for the charge
appears to be stable order-by-order. Similarly, the semiclassical Casimir model
for the fine-structure constant yields a positive value.
\end{abstract}

Recently, there have been investigations of quantum theories whose Hamiltonians
are non-Hermitian. It has been found that the energy spectra are real and
positive when these theories respect ${\cal PT}$ invariance, where ${\cal P}$
and ${\cal T}$ represent parity and time reversal. A class of quantum-mechanical
theories having this property is defined by the Hamiltonian \cite{QM}
\begin{eqnarray}
H=p^2-(ix)^N\quad(N~{\rm real}).
\label{e1}
\end{eqnarray}
For all $N\geq2$ the spectrum of $H$ is discrete, real, and positive \cite{BBM}.
Note that this theory does not respect parity invariance; thus for all $N$
(including $N=4$) the expectation value of $x$ is nonvanishing \cite{QM}. This
surprising result is a consequence of the boundary conditions \cite{ROT}.

Quantum field theories having this property have also been studied. A
generalization of Eq.~(\ref{e1}) to scalar quantum field theory is described by
the Lagrangian
\begin{eqnarray}
{\cal L}={1\over2}(\partial\phi)^2+{1\over2}m^2\phi^2-g(i\phi)^N\quad(N\geq2).
\label{e2}
\end{eqnarray}
This theory is not symmetric under $\cal P$ or $\cal T$ separately, but it is
invariant under the product ${\cal PT}$. The Hamiltonian for this theory is not
Hermitian and thus the theory is not unitary in the conventional sense. However,
there is strong evidence that the energy spectrum is real and bounded below
\cite{PARITY}. One can understand positivity heuristically in the context of the
weak-coupling expansion for the case $N=3$. The Lagrangian for this theory is
\begin{eqnarray}
{\cal L}={1\over2}(\partial\phi)^2+{1\over2}m^2\phi^2+gi\phi^3.
\label{e3}
\end{eqnarray}
In a conventional $g\phi^3$ theory the weak-coupling expansion is real, and
(apart from a possible overall factor of $g$) the Green's functions are formal
power series in $g^2$. These series are not Borel summable because they do not
alternate in sign. Nonsummability reflects the fact that the spectrum of the
underlying theory is not bounded below. However, when we replace $g$ by $ig$,
the perturbation series remains real but now alternates in sign. The
perturbation series is now summable and this suggests that the underlying theory
has a real positive spectrum.

We emphasize that replacing $g$ by $ig$ in a $\phi^3$ field theory or $g$ by
$-g$ in a $\phi^4$ field theory gives a non-Hermitian Hamiltonian. However, the
${\cal PT}$ invariance of the resulting theory is crucial and appears to
guarantee that the energy spectrum is positive.

The purpose of this note is extend these notions to quantum electrodynamics
\cite{REST}. In particular, we wish to discuss the case of massless quantum
electrodynamics and to re-examine the program of Johnson, Baker, and Willey
\cite{JBW}. In brief, the objective of their program is to find a critical value
of the coupling constant $e$ in the Lagrangian describing massless quantum
electrodynamics
\begin{eqnarray}
{\cal L}=-{1\over4}F^{\mu\nu}F_{\mu\nu}-{\bar\psi}\gamma^\mu{1\over i}
\partial_\mu\psi+e{\bar\psi}\gamma^\mu A_\mu\psi.
\label{e4}
\end{eqnarray}
The coupling constant $e$ is determined by the condition that the theory be
entirely finite. The mass shift in this theory is finite because the
unrenormalized masses are zero. Thus, the only possible infinite quantities are
associated with the three renormalization constants $Z_1$, $Z_2$, and $Z_3$.
However, to any order in powers of $e$, it is possible to find a gauge in which
$Z_1=Z_2$ is finite. Thus, the only remaining divergent quantity is $1/Z_3$.
Demanding that this be finite translates into an eigenvalue condition on the
fine structure constant
\begin{eqnarray}
\alpha={e^2\over4\pi}.
\label{e5}
\end{eqnarray}
This eigenvalue condition takes the form
\begin{eqnarray}
F_1(\alpha)=0.
\label{e6}
\end{eqnarray}

The function $F_1(\alpha)$ has been calculated to three loops (four terms) in
weak-coupling perturbation theory:
\begin{eqnarray}
F_1(\alpha)={4\over3}\left({\alpha\over4\pi}\right)+4\left({\alpha\over4\pi}
\right)^2-2\left({\alpha\over4\pi}\right)^3-46\left({\alpha\over4\pi}\right)^4
+\cdots.
\label{e7}
\end{eqnarray}
The first two terms in this series were calculated by Jost and Luttinger
\cite{JL}. Unfortunately, with just two terms the only nontrivial solution of
Eq.~(\ref{e6}) is negative, which gives an unphysical imaginary value for $e$.
In a dramatic development, Rosner \cite{ROSNER} calculated the third term in the
series. The negative sign of his result is significant because now there is a
positive root to the cubic polynomial equation obtained by truncating
$F_1(\alpha)$ after three terms:
\begin{eqnarray}
\alpha=13.872.
\label{e8}
\end{eqnarray}

Rosner's two-loop result is surprising because it is rational. His work suggests
the conjecture that all of the coefficients in the expansion of $F_1(\alpha)$
might be rational, possibly reflecting a deep symmetry of the underlying
massless theory \cite{BKZ}. This conjecture has recently gained support through
the stunning calculation of the three-loop coefficient by Gorishny and coworkers
\cite{GKLS,BK,REVIEW}. The fourth-degree equation gives one positive nontrivial
root for $\alpha$:
\begin{eqnarray}
\alpha=3.969.
\label{e9}
\end{eqnarray}
This value differs from the result in Eq.~(\ref{e8}) by a factor of $3.5$, which
suggests that this nontrivial root is unstable.

One might wonder if a stable positive root can be found by first converting the
expansion of $F_1(\alpha)$ to Pad\'e form. The $(1,1)$ Pad\'e of the Rosner
result gives no positive root at all. The $(1,2)$ Pad\'e of the four-term
series gives one positive root, $\alpha=0.814$, and the $(2,1)$ Pad\'e gives
$\alpha=0.545$. There seems to be no sensible pattern to these numerical
results.

The results regarding ${\cal PT}$-symmetric non-Hermitian quantum field theories
are intriguing because they suggest that it is possible to formulate a new kind
of electrodynamics. Instead of coupling the $A$ field to a vector current, why
not couple this field to an axial-vector current? Of course, this coupling
breaks parity symmetry. Therefore, we also replace $e$ by $ie$, thereby breaking
time-reversal invariance as well! The resulting ${\cal PT}$-symmetric, massless
Lagrangian is
\begin{eqnarray}
{\cal L}=-{1\over4}F^{\mu\nu}F_{\mu\nu}-{1\over2}\psi\gamma^0\gamma^\mu{1\over
i}\partial_\mu\psi+e{1\over2}\psi\gamma^0\gamma^5\gamma^\mu A_\mu\psi.
\label{e10}
\end{eqnarray}
We conjecture on the basis of our experience with scalar theories that the
spectrum of this theory is physically acceptable in that it is bounded below.

Our conventions in Eq.~(\ref{e10}) are as follows: $\gamma^0$ is antisymmetric
and pure imaginary, $\gamma^0\gamma^\mu$ is symmetric and real, $\gamma^5=
\gamma^0\gamma^1\gamma^2\gamma^3$ is antisymmetric and real, and $(\gamma^5)^2=
-1$. The fermion field $\psi$ is expected to be complex, as are the operators
$x$ and $p$ in Eq.~(\ref{e1}) and $\phi$ in Eqs.~(\ref{e2}) and (\ref{e3}).

Like the conventional electrodynamics described by the Lagrangian (\ref{e4}),
${\cal L}$ in (\ref{e10}) possesses gauge invariance; ${\cal L}$ is invariant
under the replacements
\begin{eqnarray}
A^\mu\to A^\mu+\partial^\mu\Lambda,\qquad\psi\to e^{-ie\gamma^5\Lambda}\psi.
\label{e11}
\end{eqnarray}
Note that this gauge transformation on the fermion field is not a phase
transformation when $e$ is real; rather it changes the scale of $\psi$. However,
the bilinear forms in the fermion field in the Lagrangian and in the
energy-momentum tensor are all invariant.

Apart from a possible overall factor of $e\gamma^5$ in some of the Green's
functions, the Feynman rules for the Lagrangian (\ref{e10}) give precisely the
same weak-coupling expansion as in the conventional massless quantum
electrodynamics (\ref{e4}), except that $\alpha$ is now replaced by $-\alpha$.
Thus, in this new and peculiar theory of quantum electrodynamics, the expansion
of $F_1(\alpha)$ becomes
\begin{eqnarray}
F_1(\alpha)=-{4\over3}\left({\alpha\over4\pi}\right)+4\left({\alpha\over4\pi}
\right)^2+2\left({\alpha\over4\pi}\right)^3-46\left({\alpha\over4\pi}\right)^4
+\cdots.
\label{e12}
\end{eqnarray}
Now, we find that there {\it is} a nontrivial positive value $\alpha_2$
satisfying the condition (\ref{e6}) when only the first two terms are retained:
\begin{eqnarray}
\alpha_2=4.189.
\label{e13}
\end{eqnarray}
If the first three terms are retained, the (unique) positive root is
\begin{eqnarray}
\alpha_3=3.657,
\label{e14}
\end{eqnarray}
which differs from $\alpha_2$ by 12\%. We feel that since $\alpha_2$ is
determined in effect from a $(2,0)$ Pad\'e it is more reasonable to convert the
three-term series to a $(2,1)$ Pad\'e and then to find the root. The slightly
different result is now
\begin{eqnarray}
\alpha_3=3.590.
\label{e15}
\end{eqnarray}
The natural continuation of this process is to calculate the $(3,1)$ Pad\'e of
the four-term series. The result is
\begin{eqnarray}
\alpha_4=4.110.
\label{e16}
\end{eqnarray}
(This is the only Pad\'e that gives a stable positive root.)

Note that the sequence of roots $\alpha_2$, $\alpha_3$, $\alpha_4$ is remarkably
stable. It would be extremely interesting to calculate the roots of the $(3,2)$,
$(4,2)$, $(4,3)$, $\ldots$, Pad\'es.

We conclude this note with a related observation. Recall that Casimir proposed a
model for determining the charge of the electron. In this model the Coulomb
repulsion of a compact charge distribution is balanced by an attractive
zero-point energy \cite{CASIMIR}. Unfortunately, although the Casimir force for
parallel plates is attractive, in a landmark paper Boyer showed that it is
repulsive for a perfectly conducting spherical shell \cite{BOYER,OTHERS}, and
thus no balance of forces is possible. However, with ${\cal PT}$-symmetric
quantum electrodynamics such a balance is achievable.

In the absence of radiative corrections, the Casimir or zero-point energy of a
perfectly conducting spherical shell of radius $a$ is
\begin{equation}
E_{\rm Casimir}={0.09235\over2a}\hbar c.
\label{e17}
\end{equation}
This energy results from fluctuations of the electromagnetic field inside and
outside the shell. This result will be unchanged in the new theory if the
boundary conditions are unaltered because the energy is independent of the
coupling to the fermion. But now, rather than a Coulomb repulsion, we have an
attraction because in effect $e\to ie$. If a charge $e$ is uniformly distributed
over a spherical shell of radius $a$, that attractive energy is
\begin{equation}
E_{\rm Coulomb}=-{1\over8\pi}{e^2\over a}.
\label{e18}
\end{equation}
Thus, stability is achieved if the two energies cancel:
\begin{equation}
E_{\rm Casimir}+E_{\rm Coulomb}=0.
\label{e19}
\end{equation}
This implies a real, positive value for the fine structure constant:
\begin{equation}
\alpha={e^2\over4\pi\hbar c}=0.09235.
\end{equation}
This is an order of magnitude larger than the physical value $1\over137$, and 40
times smaller than the value found above for a finite quantum electrodynamics.
But what is significant here is that a positive solution for $\alpha$ actually
exists.

\section* {ACKNOWLEDGMENTS}
\label{s1}

We thank A.~Kataev and D.~Broadhurst for helpful communications and we are
grateful to the U.S.~Department of Energy for financial support.

\end{document}